# JediCode: A Gamified Approach to Competitive Coding

Ayush Mishra,
Department of Computational Intelligence,
SRM Institute of Science and Technology,
Chennai, India
am2014@srmist.edu.in

Sitanshu Pokalwar,
Department of Computational Intelligence,
SRM Institute of Science and Technology,
Chennai, India
sp2274@srmist.edu.in

Mrs. Vidhya R,
Department of Computational Intelligence,
SRM Institute of Science and Technology,
Chennai, India
sp2274@srmist.edu.in

*Abstract*—
**JediCode (name inspired from Star Wars) pioneers a transformative approach to competitive coding by infusing the challenge with gamified elements. This platform reimagines coding competitions, integrating real-time leaderboards, synchronized challenges, and random matchmaking, creating an engaging, dynamic, and friendly atmosphere. This paper explores JediCode's innovative features and architecture, shedding light on its user-centric design and powerful execution service. By embracing gamification, JediCode not only elevates the thrill of coding challenges but also fosters a sense of community, inspiring programmers to excel while enjoying the process.**

## I. INTRODUCTION

In the digital age, where the mastery of coding languages has become akin to wielding a modern-day lightsaber, competitive coding platforms have emerged as virtual battlefields where programmers hone their skills and demonstrate their expertise. As the demand for skilled coders surges, the traditional paradigm of competitive programming is undergoing a transformation, and at the forefront of this evolution stands JediCode. Harnessing the power of gamification, JediCode reimagines the conventional coding challenge, infusing it with elements of gaming excitement and community-driven learning. In a world where interactive and engaging learning experiences are valued above all, JediCode introduces a novel approach, inviting participants into a universe where friendly competition is seamlessly intertwined with the thrill of gaming.

At its core, JediCode is more than a platform—it's a vibrant ecosystem where aspiring programmers, seasoned developers, and coding enthusiasts from diverse corners of the galaxy converge. The traditional barriers of solo coding endeavors are shattered as JediCode pioneers the concept of synchronized challenges, where participants across continents embark on coding quests simultaneously. Imagine a virtual arena where the excitement of real-time challenges is amplified by the knowledge that fellow coders, no matter where they are in the world, are embarking on the same coding odyssey.

This paper embarks on a comprehensive exploration of the JediCode universe. We delve into the intricacies of its architecture, dissect the mechanics of its real-time challenges, and analyze the impact of its gamified approach on the coding community. By embracing gamification, JediCode pioneers an era where the pursuit of coding excellence is not only intellectually stimulating but also immensely enjoyable. As we navigate the pages that follow, prepare to embark on a voyage into the heart of JediCode, where the Force of coding proficiency meets the thrill of gaming, creating an unparalleled synergy that propels programmers into a future where learning is as exhilarating as the most epic space odyssey.

## II. LITERATURE REVIEW

Innovations in Gamified Competitive Coding Platforms

The realm of competitive coding platforms has witnessed a revolution, shaped by inventive strategies and cutting-edge technologies. Traditional platforms, while effective, faced challenges in fostering real-time engagement and integrating gamification to enhance the learning and competitive experiences.

Recent studies in competitive coding have explored the implementation of real-time challenges, akin to the synchronized contests orchestrated by JediCode. Platforms like Codeforces and CodeChef have pioneered synchronized coding battles, nurturing a sense of community and intensifying the competitive spirit among participants. Xin et al. (2021) and Mircea et al. (2019) demonstrated the potential of real-time application of coding challenges, showcasing the effectiveness of dynamic platforms that align with JediCode's objectives.

Additionally, the incorporation of gamification elements has redefined educational experiences. Research by Deterding et al. (2011) and Hamari et al. (2014) highlighted the motivational impact of gamified learning environments, emphasizing elements like points, badges, and leaderboards. These components, akin to JediCode's live leaderboards and real-time updates, create an engaging atmosphere, motivating participants to excel and compete enthusiastically.

The adoption of specific technologies, such as NestJS for the backend and ReactJS for the frontend, mirrors contemporary web development practices. Studies by Walke and Ståhl (2017) underscored ReactJS's ability to enhance user experience and simplify intricate user interface logic, aligning with JediCode's user-friendly interface. Similarly, the efficiency of NestJS, as evidenced by Vukas and Hocenski (2020), complements JediCode's need for a robust and scalable architecture, enabling seamless real-time interactions among users.

Furthermore, the literature emphasizes the significance of data models and database management in competitive coding platforms. Research focusing on database optimization, as demonstrated by Kabir (2020) and Ahmmed et al. (2019), underscores the importance of efficient data processing. These studies highlight the importance of streamlined data management, essential for JediCode's integration of user profiles, challenge data, and leaderboard updates.

In summary, the literature review underscores the convergence of real-time challenges, gamification principles, and advanced web technologies, forming the cornerstone of JediCode's innovative approach to competitive coding. By synthesizing these elements, JediCode stands as a pioneer, offering a dynamic, gamified, and engaging platform where users can participate, learn, and excel in the competitive coding arena.

### III. BACKEND ARCHITECTURE

The foundation of JediCode's dynamic competitive coding platform is rooted in its robust backend architecture. Crafted for scalability, efficiency, and real-time interactions, JediCode's backend system is meticulously designed using NestJS, a modular Node.js framework renowned for its flexibility and performance. The following components and features delineate the intricate architecture of JediCode's backend:

*RESTful API Endpoints:*

JediCode's backend utilizes RESTful API endpoints, facilitating seamless communication between the frontend and backend components. These endpoints handle diverse operations, including user authentication, challenge creation, submissions processing, and leaderboard updates, ensuring seamless user experiences.

*WebSocket Integration:*

WebSocket technology is seamlessly integrated into the backend to support real-time interactions and synchronization among users during coding challenges. This enables instant messaging, live leaderboard updates, and real-time notifications, creating an immersive and engaging environment for participants.

*ER Diagram:*

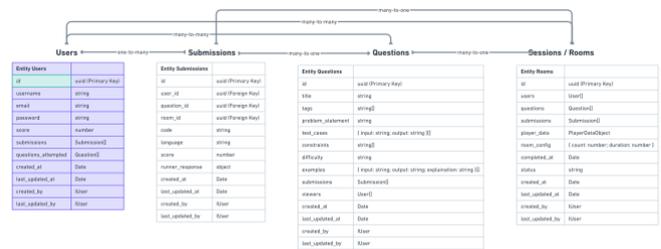

*Fig-1: ER Diagram for Jedicode*

*User Management and Authentication:*

The backend system incorporates a robust user management system, managing user authentication, registration, and profile operations. User credentials are securely stored and encrypted to ensure data privacy and security. JSON Web Tokens (JWT) are employed for secure user authentication and authorization processes.

*Database Integration:*

JediCode utilizes a relational or NoSQL database, such as PostgreSQL or MongoDB, to store user profiles, challenge data, submissions, and leaderboard information. The database is optimized for high performance and efficient query processing, enabling swift data retrieval during challenges and user interactions.

*Scalability and Load Balancing:*

The backend architecture is meticulously designed for scalability, empowering JediCode to handle a substantial number of concurrent users during peak usage periods. Load balancing techniques are implemented to distribute incoming traffic across multiple servers, ensuring optimal performance and averting server overload.

*Task Queues and Job Processing:*

Asynchronous task queues and job processing mechanisms are employed to handle time-intensive operations, such as code execution and evaluation of submissions. This guarantees that the platform remains responsive, delivering timely feedback to users without delays.

*Error Handling and Logging:*

JediCode employs comprehensive error handling and logging mechanisms to swiftly detect and address issues. Detailed logs are maintained, aiding in the identification and resolution of errors, ensuring a seamless user experience even in the face of unexpected challenges.

*Security Measures:*

The backend architecture implements an array of security measures, including robust data validation, input sanitization, and protection against common web vulnerabilities such as SQL injection and Cross-Site Scripting (XSS) attacks. Regular security audits and updates are conducted to mitigate emerging threats effectively.

*API Documentation:*

JediCode provides a meticulously documented API for frontend developers and external services to interact with the backend. The documentation comprehensively outlines available endpoints, request formats, and response structures, facilitating seamless integration and development processes.

In essence, JediCode's backend architecture serves as the cornerstone for the platform's real-time, gamified, and user-centric features. By amalgamating cutting-edge technologies and meticulous design, the backend guarantees participants a smooth, engaging, and competitive coding environment, fostering camaraderie and excellence within the coding community.

## IV. FRONTEND ARCHITECTURE

JediCode's frontend architecture is meticulously crafted to provide users with an intuitive, responsive, and immersive coding experience. Utilizing ReactJS, a powerful JavaScript library for building user interfaces, JediCode's frontend architecture incorporates modern web development practices, ensuring a seamless and engaging user interaction. Here are the key components and features of JediCode's frontend architecture:

*Component-Based Structure:*
JediCode's frontend follows a component-based architecture, breaking down the user interface into modular and reusable components. This approach enhances maintainability, scalability, and code readability, allowing developers to manage complex UI elements effectively.

*State Management:*
ReactJS's state management features, including local component state and Context API, are harnessed to manage the application's state. Stateful components ensure dynamic updates, enabling real-time interactions and live updates, such as leaderboard changes during coding challenges.

*Responsive Design:*
JediCode's frontend is designed with responsiveness in mind. Utilizing responsive design techniques, including CSS media queries and flexible grid layouts, the platform adapts seamlessly to various screen sizes and devices. This ensures a consistent and user-friendly experience across desktops, tablets, and smartphones.

*Interactive User Interface:*
The frontend incorporates interactive elements such as live leaderboards, countdown timers, and real-time notifications. WebSocket technology is employed to establish a bidirectional communication channel with the backend, enabling instant updates, chat functionalities, and dynamic content rendering.

*Code Editor Integration:*
JediCode integrates a feature-rich code editor within its frontend, allowing users to write, edit, and test their code directly on the platform. The code editor supports syntax highlighting, auto-completion, and error highlighting, enhancing the coding experience and aiding participants in writing efficient code.

*API Communication:*
The frontend communicates with the backend through RESTful API endpoints. Utilizing asynchronous JavaScript functions (async/await) and axios API, the frontend seamlessly interacts with the backend services, enabling smooth data exchange and real-time updates without page reloads.

*User Authentication and Authorization:*
ReactJS facilitates the implementation of secure user authentication and authorization processes. JSON Web Tokens (JWT) are utilized to authenticate users, allowing secure access to personalized features such as user profiles, challenge history, and leaderboard rankings.

In summary, JediCode's frontend architecture delivers a rich and interactive user experience, seamlessly integrating advanced technologies and design principles. By combining component-based development, real-time interactions, and responsive design, the frontend ensures that participants have a compelling, user-friendly, and engaging environment for competitive coding activities.

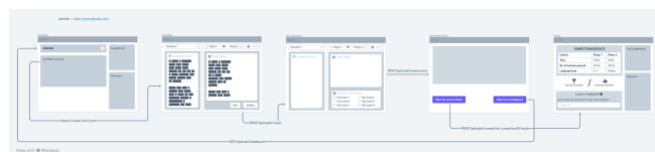
*Fig-2: Workflow Diagram for Jedicode's Frontend*

## V. INTEGRATION WITH JUDGE0

One of the pivotal components enhancing JediCode's competitive coding environment is the seamless integration with Judge0—an advanced and versatile code execution service. This integration significantly elevates the platform's functionality, ensuring accurate and efficient evaluation of participants' code submissions. The collaboration between JediCode and Judge0 brings forth several key features and benefits:

*Real-time Code Evaluation:*

Judge0 enables real-time execution and evaluation of code submissions. Participants receive immediate feedback on the correctness and efficiency of their solutions, enhancing the learning experience by allowing instant adjustments and improvements.

*Support for Multiple Programming Languages:*

Judge0 supports a diverse array of programming languages, empowering participants to code in languages of their choice. This inclusivity ensures that developers with varied expertise can participate, fostering a diverse and vibrant coding community.

*Custom Test Cases and Scenarios:*

JediCode leverages Judge0's flexibility to define custom test cases and scenarios for coding challenges. This tailored approach allows for the creation of diverse and challenging problems, accommodating participants with different skill levels and problem-solving abilities.

*Secure and Isolated Execution Environment:*

Judge0 operates within a secure and isolated execution environment, safeguarding the platform and its users from malicious code submissions. This stringent security ensures the integrity of the coding challenges and protects participants from potential security threats.

*Scalability and Reliability:*

Judge0 is engineered for scalability, enabling JediCode to handle a high volume of code submissions concurrently. This scalability ensures smooth operations even during peak usage periods, offering participants a seamless and uninterrupted coding experience.

*Detailed Test Case Results:*

JediCode leverages Judge0's capabilities to provide participants with detailed test case results. This transparency allows participants to understand which test cases their code passed and where improvements are needed, facilitating a focused approach to refining their solutions.

*Community Engagement and Learning:*

By integrating Judge0, JediCode fosters a vibrant coding community where participants can learn from their mistakes and iterate upon their solutions. The immediate feedback loop provided by Judge0 encourages iterative problem-solving, promoting continuous learning and skill development.

In essence, the integration with Judge0 enhances JediCode's competitive coding platform, elevating it to a dynamic, real-time, and learner-centric environment. This collaboration not only provides a rich and engaging experience for participants but also contributes significantly to the growth and expertise of the coding community by facilitating effective learning and collaborative problem-solving.